\documentclass[12pt]{article}

\usepackage{amssymb}
\usepackage{graphicx}
\usepackage{tabularx}
\usepackage{tabulary}
\usepackage{multirow}
\usepackage[nodots]{numcompress}
\usepackage{fancyhdr}

\begin{document}
\title{Forensic Analysis of WhatsApp Messenger \\  on Android Smartphones}
\author{Cosimo Anglano \\
DiSIT - Computer Science Institute,\\ Universit\'{a} del Piemonte Orientale, Alessandria (Italy)\\
email:cosimo.anglano@uniupo.it}
\date{}
\maketitle
\begin{center}
	\textbf{This manuscript is the preprint of the paper \\
		\emph{Forensic Analysis of WhatsApp Messenger on Android Smartphones}, \\
		that has been published on the \emph{Digital Investigation} Journal, Vol. 11, No. 3, pp. 201--213,
		September 2014 \\
	doi:10.1016/j.diin.2014.04.003 \\
	(c) 2014. This manuscript version is made available under the CC-BY-NC-ND 4.0 license http://creativecommons.org/licenses/by-nc-nd/4.0/
	}
\end{center}
\begin{abstract}
We present the forensic analysis of the artifacts
left on Android devices by \textit{WhatsApp Messenger}, the client
of the WhatsApp instant messaging system.
We provide a complete description of all the artifacts
generated by WhatsApp Messenger, we discuss the decoding and the interpretation 
of each one of them, and we show how they can be correlated together to 
infer various types of information that cannot be obtained by considering each one of
them in isolation.

By using the results discussed in this paper, an analyst will be able
to reconstruct the list of
contacts and the chronology of the messages that have been exchanged
by users.
Furthermore, thanks to the correlation of multiple artifacts, (s)he will be able to infer 
information like when a specific contact has been
added, to recover deleted contacts and their time of deletion, to
determine which messages have been deleted, when these messages have been
exchanged, and the users that exchanged them.
\end{abstract}


\section{Introduction}
\label{intro}
The introduction of sophisticated communication services over the Internet,
allowing users to exchange textual messages, as well as audio, video,
and image files, has changed the way people interact among them.
The usage of these services, broadly named \textit{instant messaging (IM)},
has undoubtedly exploded in the past few
years, mainly thanks to the pervasiveness of smartphones, that provide
quite sophisticated IM applications.
Smartphones indeed enable users to exploit their data connection to
access IM services anywhere and anytime, 
thus eliminating the costs usually charged by mobile operators for similar
services (e.g., for SMS communication).

Given their popularity, IM services are being increasingly used not
only for legitimate activities, but also for illicit ones~\cite{unodc-2013}:
criminals may indeed use them
either to communicate with potential victims, or with other criminals
to escape interception~\cite{bellovin-wiretapping}.
Therefore, IM applications have the potential of being a very rich
source of evidentiary information in most investigations.

Among IM applications for smartphones, \textit{WhatsApp}~\cite{whatsapp} is accredited 
to be the most widespread one (reportedly~\cite{wa:400m}, it  has 
over 400 million active users that exchange, on average, more than 31 billion messages 
per day, 325 millions of which are photos~\cite{wa:300m}).
Given its recent acquisition by Facebook, it is reasonable to expect 
a further growth in its diffusion.
Therefore, the analysis of \textit{WhatsApp Messenger}, the client of 
WhatsApp that runs on smartphones, has recently raised 
the interest of the digital forensics community~\cite{thakur-master-thesis,mahajan-ijca-2013,tso-iphone}.

In this paper we deal with the forensic analysis of WhatsApp Messenger on Android
smartphones. 
Android users, indeed, arguably represent the largest part of the user base of WhatsApp: as of 
Jan. 2014, Google Playstore reports a number of downloads  
included between $100$ and $500$ millions (the lower limit having 
been already hit in  Nov. 2012), out of a population of $400$ millions
of users.
Thus, by focusing on the Android platform, we maximize the potential
investigative impact of our work. 

Several works, appeared recently in the literature~\cite{thakur-master-thesis,mahajan-ijca-2013}
deal with the same problem. However, as discussed later, these works
are limited in scope, as they focus on only the reconstruction
of the chronology of exchanged messages, and neglect other important
artifacts that, instead, are considered in our work.

More precisely, the contributions of this paper can be summarized
as follows:
\begin{itemize}
\item we discuss the decoding and the interpretation of all the artifacts and data
generated by WhatsApp Messenger on Android devices;
\item we show how these artifacts can be correlated together to 
infer various types of information that cannot be obtained by considering each one of
them in isolation, such as when a contact has been added to or deleted from
the contacts database, whether a message
has been actually delivered to its destination after having been sent
or has been deleted,
if a user joined or left a group
chat before or after a given time, when a given user has been added to the
list of contacts, etc..
\end{itemize}

The rest of the paper is organized as follows. In Sec.~\ref{sec:related}
we review existing work, while in Sec.~\ref{sec:methodology} we describe
the methodology and the tools we use in our study.
Then, in Sec.~\ref{sec:analysis} we discuss the forensic analysis of
WhatsApp Messenger and, finally, in Sec.~\ref{sec:conclusions} we
conclude the paper and outline future research work.

\section{Related works}
\label{sec:related}
The forensic analysis of IM applications on smartphones
has been the subject of various works published in the
literature.

Compared with existing works, however, our contribution (a) has a wider scope, as
it considers all the artifacts generated by WhatsApp Messenger (namely,
the database of contacts, the log files, the avatar pictures, and the preference files),
(b) presents a more thorough and complete analysis of these artifacts,
and (c) explains how these artifacts can be correlated to deduce
various type of information having an evidentiary value, such as whether a message
has been actually delivered to its destination after having been sent, 
if a user joined or left a group
chat before or after a given time, and when a given user has been added to the
list of contacts.

\cite{iforensics} focus on the forensic analysis of
three IM applications (namely AIM, Yahoo! Messenger, and Google
Talk) on the iOS platform.
Their work differs from ours for both the IM applications
and the smartphone platform it considers.

\cite{social-im-forensics} focus on the analysis of 
several IM applications (including WhatsApp Messenger) on various smartphone
platforms, including Android, with the aim of identifying the encryption
algorithms used by them.
Their work, unlike ours, does not deal with the identification,
analysis, and correlation of all the artifacts generated
by WhatsApp Messenger.

\cite{tso-iphone} focus on the analysis of iTunes backups for
iOS devices with the aim of identifying the artifacts left by
various social network applications, including WhatsApp Messenger.
Their work differs from ours because of its focus on iTunes and iOS,
and because only the chat database of WhatsApp is considered, since
only this artifact is included into an iTune backup.
Furthermore, the information stored into the chat database is analyzed
only in part.

The works of \cite{thakur-master-thesis} and \cite{mahajan-ijca-2013} are similar
to ours, since they both focus on the
forensic analysis of WhatsApp Messenger on Android.
However, these works focus mainly on the forensic acquisition
of the artifacts left by WhatsApp Messenger, and deal with
their analysis only in part (they limit their study to the
chat database, and analyze it only partially).
Similar considerations apply to the \textit{WhatsApp Xtract} tool~\cite{sangiacomo-wa-xtract},
that extracts some of the information stored into the chat database (and, possibly,
in the contacts database), without however providing any description of
how these databases are parsed.

\section{Analysis methodology and tools}
\label{sec:methodology}
The study described in this paper has been performed by
carrying out a set of controlled experiments, each one referring to a specific
usage scenario (one-to-one communication, group communication, multimedia message exchange, etc.), 
during which typical user interactions have taken place. 
After each experiment, the memory of the sending and receiving devices
has been examined
in order to identify, extract, and analyze the data generated 
by WhatsApp Messenger in that experiment.

As discussed in Sec.~\ref{sec:analysis}, most of the files
generated by WhatsApp Messenger are stored into an area of the internal device memory 
that is normally inaccessible to users.
To access this area, suitable commercial tools may be used~\cite{ufed,xry,oxygen}
but, unfortunately, we did not have access to them.
Open-source software-based tools are also available~\cite{hoog-book,android-acquisition},
but we consider problematic their use in our study for the following reasons:
\begin{itemize}
\item they may alter the contents of the memory, thus overwriting pieces
of information: 
while this can be considered acceptable in a real-world investigation 
when there is no other alternative, we believe that 
it should be avoided in a study like the one presented in this paper,
as modified or incomplete data may yield to incorrect conclusions;
\item they are device-specific: this would prevent a third party to replicate
the experiments to validate our findings, unless the same device model and the
same software acquisition tool are used.
\end{itemize}

For these reasons, in our study we adopted a different approach,
in which we use software-emulated Android devices in place of physical ones.
In particular, we use the YouWave virtualization
platform~\cite{youwave} that is able to faithfully
emulate the behavior of a complete Android device.
YouWave implements the internal device memory as a VirtualBox storage file~\cite{virtualbox},
whose format is documented and, therefore, can be parsed by a suitable tool
to extract the files stored inside it.
In this way, the acquisition of the internal memory of the device is greatly simplified, as
it reduces to inspect the content of this file.

In order to ensure the soundness of our approach, we have made
tests in which the behavior of, and the data generated by, WhatsApp
Messenger running on YouWave have been compared against those produced
when it runs on real smartphones.
These tests have been performed either indirectly, by
comparing the data found in the inaccessible memory
area of YouWave against those documented in the literature\cite{thakur-master-thesis,mahajan-ijca-2013}, or directly, by comparing the data stored on the emulated SD memory
card against those generated on a real smartphone.
The results of our tests indicate that, from the perspective of WhatsApp
Messenger, YouWave and  a real smartphone behave the same way.

Our experimental test-bed consists thus into a set
of YouWave virtual machines, namely one for each device involved
in the experiments, running Android v. 4.0.4.
On each one of these machines we install and use
WhatsApp Messenger v. 2.11.
In each experiment, we assign a role to each virtual device
(e.g. sender or recipient of a message, group chat leader,
etc.), and use it to carry out the corresponding activities.
Then, at the end of the experiment, we suspend the virtual
device, parse the file implementing the
corresponding internal memory (named \textsf{youwave\_vm01.vdi})
by means of \emph{FTK Imager}(v. 3.1)~\cite{ftk-imager},
and extract the files where WhatsApp Messenger stores the data it generates.~\footnote{
The only exception we make to the above methodology is the use
of a physical smartphone to generate messages carrying geolocation
coordinates, as the Android Location Services, used by WhatsApp Messenger
to obtain the coordinates of the current location of the device,
are not available on YouWave because of its lack of
a GPS receiver. In this case, access to the relevant data
is achieved by using the backup mechanisms described in
Sec.~\ref{sec:chat-db}.}
These files are then examined by means of suitable tools.
In particular, we use \textsl{SqliteMan}~\cite{sqliteman}
to examine the databases maintained by WhatsApp Messenger
(as discussed later, they are SQLite v.3 databases~\cite{sqlite3}),
and \textsf{notepad++}~\cite{notepad++} to examine
textual files.

By proceeding as exposed above, (a) we are able to avoid the risks of contamination and
of an incomplete acquisition of the data stored in the memory of the device, (b) we ensure 
repeatability of experiments, as their outcomes do not depend on the availability of a 
specific software or hardware memory acquisition tool or smartphone model, 
(c) we obtain a high degree of
controllability of experiments, as we may suspend and resume at will the virtual device to
perform acquisition while a given experiment is being carried out
and, last but not less important, (d) we reduce the costs of the study, since neither real smartphones nor commercial memory acquisition tools are 
necessary to carry out the experiments.

\section{Forensic analysis of WhatsApp Messenger}
\label{sec:analysis}
WhatsApp provides its users with various forms of communications, namely
user-to-user communications, broadcast messages, and group chats.
When communicating, users may exchange plain text messages, as well as
multimedia files (containing images, audio, and video), contact cards, 
and geolocation information.

Each user is associated with its \textit{profile}, a set of information that includes
his/her \textit{WhatsApp name}, \textit{status line}, and \textit{avatar}
(a graphic file, typically a picture).
The profile of each user is stored on a central system,
from which it is downloaded by other WhatsApp users that include that user
in their contacts. 
The central systems provides also other services, like user registration,
authentication, and message relay. 

As reported in \cite{thakur-master-thesis},
the artifacts generated by WhatsApp Messenger on an Android device 
are stored into a set of files, whose name,
location, and contents are listed in Table~\ref{tab:wa-artifacts-location}.

\begin{table}[hbtp]
\begin{center}
\begin{small}
    \begin{tabulary}{\linewidth}{|C|L|L|L|}
    
        \hline
        \textbf{Row \#} & \textbf{Content} & \textbf{Directory} & \textbf{File}\\ \hline\hline
         1 & contacts database   & /data/data/ com.whatsapp/databases & \textsf{wa.db} (SQLite v.3)\\ \hline
         2 & chat database & /data/data/ com.whatsapp/databases & \textsf{msgstore.db} (SQLite v.3)\\ \hline
         3 & backups of the chat database & /mnt/sdcard/ Whatsapp/Databases & \textsf{msgstore.db.crypt} \textsf{msgstore-$<$date$>$.crypt} \\ \hline
         4 & avatars of contacts  & /data/data/ com.whatsapp/files/ Avatars & \textsf{UID.j}, where \textsf{UID} is the identifier of the contact\\ \hline
         5 & copies of contacts avatars & /mnt/sdcard/  WhatsApp/ProfilePictures &  \textsf{UID.j}, where \textsf{UID} is the identifier of the contact\\ \hline
         6 & log files   & /data/data/ com.whatsapp/files/ Logs &  \textsf{whatsapp.log},  \textsf{whatsapp-$<$date$>$.log} \\ \hline
		 7 & received files  & /mnt/sdcard/ Whatsapp/Media & various files  \\  \hline
		 8 & sent files  & /mnt/sdcard/ Whatsapp/Media/Sent &  various files \\ \hline
		 9 & user settings and preferences & /data/data/ comm.whatsapp/files & various files \\ \hline		\end{tabulary}
\end{small}
\end{center}
\caption{WhatsApp Messenger artifacts.}
\label{tab:wa-artifacts-location}
\end{table}

In the rest of this section we discuss how the above artifacts can be
analyzed and correlated to ascertain various types of information:
we start with contact information (Sec.~\ref{sec:contacts}), 
we continue with exchanged messages (Sec.~\ref{sec:messages-analysis}),
and we end with application settings and user
preferences (Sec.~\ref{sec:settings-and-preferences}).
\subsection{Analysis of contact information}
\label{sec:contacts}
The evidentiary value of contact information is notorious,
as it allows an investigator to determine who
the user was in contact with.

In this section we first describe the information
that are stored in the contacts database, 
and then we discuss how this information
can be analyzed to determine (a) the list of contacts,
(b) when a contact has been added to the
database, (c) whether and when a given contact has been blocked and,
finally, we show how deleted contacts can be dealt with.
\subsubsection{Retrieving contact information}
\label{sec:getting-contacts}
The contacts database \textsf{wa.db} contains
three tables, namely \textsf{wa\_contacts}, that stores a record
for each contact, \textsf{android\_metadata}, and
\textsf{sqlite\_sequence}, both storing housekeeping information
having no evidentiary value.

The structure of the records in \textsf{wa\_contacts} is shown in Table~\ref{wa-table}, where
we distinguish the fields containing data obtained from the
WhatsApp system (and, as such, having potential evidentiary value),
from those storing data extracted from the phonebook
of the device (that, being set by the user and not by WhatsApp,
are not pertinent to our work).
\begin{table}[hbtp]
\begin{center}
\begin{small}

    \begin{tabularx}{\linewidth}{|l|X|}
        \hline
        \multicolumn{2}{|c|}{\textbf{Data coming from the WhatsApp system}} \\ \hline
        \textbf{Field name}  & \textbf{Meaning} \\ \hline
        \_id	& sequence number of  the record (set by SQLite) \\ \hline
        jid     & WhatsApp ID of the contact (a string structured as
'\textsf{x@s.whatsapp.net}', where '\textsf{x}' is the phone number
of the contact) \\ \hline
        is\_whatsapp\_user  &  contains '1' if the contact corresponds to an actual WhatsApp user, '0' otherwise \\ \hline
        unseen\_msg\_count	& number of messages sent by this contact that have been received, but still have to be read\\ \hline
        photo\_ts			& unknown, always set to '0'\\ \hline
        thumb\_ts			& Unix epoch time (10 digits) indicating when the contact has set his/her current avatar picture\\ \hline
        photo\_id\_timestamp &  Unix millisecond epoch time (13 digits) indicating when the current avatar picture of the contact has been downloaded locally\\ \hline
        wa\_name			& WhatsApp name of the contact (as set in his/her profile)\\ \hline
        status		& status line of the contact (as set in his/her profile) \\ \hline
        sort\_name 	        &  name of the contact used in sorting operations \\ \hline \hline  
        \multicolumn{2}{|c|}{\textbf{Data coming from from the phonebook of the device}} \\ \hline
        \textbf{Field name}  & \textbf{Meaning} \\ \hline
        number		 		& phone number associated to the contact \\ \hline
        raw\_contact\_id	& sequence number of the contact   \\ \hline
        display\_name		& display name of the contact \\ \hline
        phone\_type			& type of the phone  \\ \hline
        phone\_label		& label associated to the phone number  \\ \hline
        given\_name			& given name of the user  \\ \hline
        family\_name		& family name of the user \\ \hline
    \end{tabularx} 
\end{small}
\end{center}
\caption{Structure of the \textit{wa\_contacts} table}
\label{wa-table}
\end{table}

As can be observed from this table, each record stores the
WhatsApp ID (field \textsf{jid}) of the contact, a string structured as
'\textsf{x@s.whatsapp.net}', where '\textsf{x}' is the phone number
of that contact (for the sake of readability, in the following
we indicate users by means of their phone numbers instead of their complete
WhatsApp IDs).
Furthermore, each record stores 
the profile name (field  \textsf{wa\_name}),
and the status string (field \textsf{status}) of the corresponding contact.
Field \textsf{is\_whatsapp\_user} is instead used to differentiate
actual WhatsApp users from unreal ones: WhatsApp Messenger indeed adds to the contact
database a record for each phone number found in the phonebook of the device, even if the
corresponding user is not registered with the WhatsApp system.

Avatar pictures may have evidentiary value as well: they can be indeed used to link a 
WhatsApp account to the real identity of the person  using it (for instance, if the 
avatar displays the face of the user, or any location or item that can be uniquely 
associated with that person).
The avatar picture of a contact  \textsf{x@s.whatsapp.net} is stored, as a JPEG file
named \textsf{x@s.whatsapp.net.j}, in the directories listed in 
Table~\ref{tab:wa-artifacts-location}, rows no. 4 and 5.
The timestamps stored in the \textsf{thumb\_ts} and \textsf{photo\_id\_timestamp} field
indicate when the contacts has set his/her current avatar, and when
that avatar has been downloaded locally, respectively. 

\subsubsection{Determining when a contact has been added}
\label{sec:adding-contact}
In some investigations, it may  be necessary to determine when a given
user has been added to the contacts database~\footnote{User contacts are automatically 
added to the contacts database by 
WhatsApp Messenger that -- each time is started or when the user
starts a new conversation -- inspects the phonebook of the device and
adds all the phone numbers that are not stored there yet.}.
This information is not stored in the \textsf{wa\_contacts} table, but can
be deduced from the analysis of the
log files generated by WhatsApp Messenger (that are located in the
directory listed in Table~\ref{tab:wa-artifacts-location}, row no. 6).

When a contact is added to the \textsf{wa.db} database, 
WhatsApp Messenger logs several events that are tagged with their
time of occurrence and with the WhatsApp ID of the involved user.
\begin{figure}[hbt]
\centering
\includegraphics[scale=0.75]{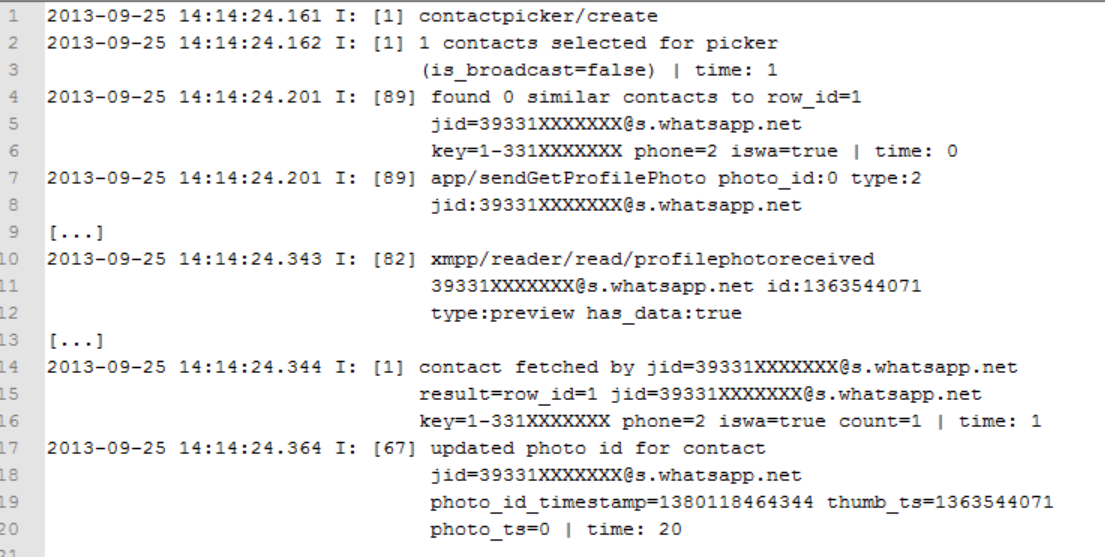}
\caption{Events logged when a user is added to the
\textsf{contacts} table(phonenumber redacted to ensure
the privacy of the owner). Long lines have been split to ease readability.}
\label{fig:log-added-contact}
\end{figure}

Examples of these events, corresponding to the addition
of user \textsl{39331xxxxxxx}, are reported in Fig.~\ref{fig:log-added-contact},
from which we note that the following events are logged each time
a new user is added:
(a) the discovery that the user is not present yet in the contacts database
(line no. 4), (b) the queries to the central system to fetch various
information about the contact (lines no. 7,10, and 14), and (c) the
completion of the download of the corresponding avatar picture
(line no. 17).
From these events, we can determine when the user has
been added to the contacts database (on Sept. 25, 2013 at 14:14:24,
in our example).
\subsubsection{Dealing with blocked contacts}
\label{sec:blocked-contacts}
WhatsApp Messenger enables the user to block anyone of his/her contacts,
thus preventing any communication with him/her until the block
is removed.
In an investigation it can be important to determine whether a contact
was blocked or not at a given time, in order to confirm or to exclude
the reception of a message sent at that time. 

The information concerning blocked users is stored neither in the
contacts database, nor elsewhere on the memory of the device
(we conjecture that the list of blocked contacts 
is stored on the WhatsApp central system, since when
the blocking is taking place, WhatsApp Messenger exchanges messages 
with it).
Blocked  users can be however identified, under some circumstances,
by analyzing log files.

\begin{figure}[hbt]
\centering
\includegraphics[scale=0.6]{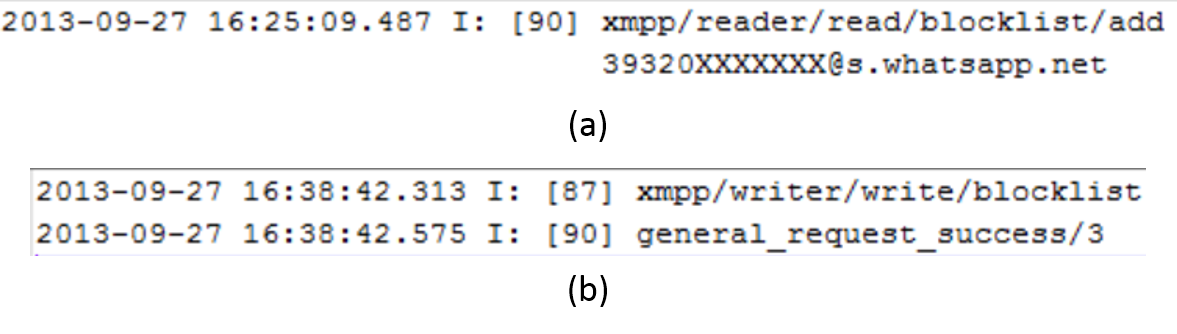}
\caption{Events in the log file corresponding to (a) the blocking, (b) the unblocking of 
user \textsl{39320xxxxxxx}.}
\label{fig:log-contact-block-unblock}
\end{figure}

When a contact is blocked, an event -- reporting the WhatsApp ID of that contact
and the time of occurrence of the  operation -- is indeed
recorded into the log file (see Fig.~\ref{fig:log-contact-block-unblock}(a)).
Unfortunately, when a contact is unblocked, the event that is
logged (Fig.~\ref{fig:log-contact-block-unblock}(b)) 
does not report the WhatsApp IDs of the involved contact, 
and it is cumulative (i.e., it may refer to 
a set of contacts being unblocked simultaneously).

Thus, it is always possible to determine whether and when a given user \textsf{X} has been
blocked, but whether it is still blocked
at a given time can be ascertained only if either (a) no unblocking events are
recorded in the log file after the block operation, or (b)
an unblocking event is present, but only user \textsf{X} was blocked
at that time.
It follows that if several users are blocked and one (or more) unblocking
events are logged, it is not possible to tell which users are still blocked,
and which ones have been instead unblocked.
It is worth pointing out that the above inferences can be made only
if the log files reporting blocking and unblocking events are still
available (i.e., they have not been deleted  by WhatsApp Messenger
to create room for newer  ones).

As a final consideration, we note that no information whatsoever is stored on the
side of the contact that gets blocked, so it is not possible to tell whether
the user of the device under analysis has been blocked or not by anyone of
his/her contacts. 
\subsubsection{Dealing with deleted contacts}
\label{sec:deleted-contacts}
In the attempt to hide past interactions, the user may delete a contact,
thus causing the removal of the corresponding record from the
\textsf{wa\_contacts} table.

In some cases (notably, if the SQLite engine has not vacuumed
the above table yet), it may be possible to recover deleted records by means of
suitable techniques (e.g.,~\cite{sqlite-deletion,foxit-dfrws2011}.
Our experiments, carried out by means of Oxygen Forensic SQLite Viewer~\cite{oxygen-sqlite},
indicate indeed that deleted contact records may be recovered.

However, in general, at the moment of the analysis, deleted records may have been vacuumed,
so they cannot be recovered anymore.
In these situations, it may be still possible to determine the set of
deleted contacts by first reconstructing the list of contacts that have been
added in the past (by analyzing log files as discussed in Sec.~\ref{sec:adding-contact}),
and then by comparing this list with the contents of the \textsf{wa\_contacts} table:
the contacts in the list that are not in the database
are those that have been deleted.
Note that this procedure works only if the log file reporting  the
addition of a contact of interest is still available when the analysis
is performed.

Unfortunately, by proceeding as above, it is not possible to 
determine when a given contact has been deleted,
since
deletions give rise to log events that do not reference
the WhatsApp ID of the contact being deleted.

\subsection{Analysis of exchanged messages}
\label{sec:messages-analysis}
WhatsApp Messenger stores
all the messages that have been sent or received 
into the chat database \textsf{msgstore.db} (located in
the directory listed in Table~\ref{tab:wa-artifacts-location},
row 2), whose analysis makes it possible to reconstruct
the chronology of exchanged messages, namely to determine 
when a message has been exchanged, the data it carried, the set of users 
involved in the conversation, and whether and when it has
actually been received by its recipients.

In the following we discuss each one of the above steps separately:
we start with the description of the structure of the 
chat database (Sec.~\ref{sec:chat-db}), and then we explain how to (a) reconstruct
the chat history (Sec.~\ref{sec:chat-history}), (b) determine
and extract the content of a message (Sec.~\ref{sec:conversation-content}),
(c) determine the status of a message (Sec.~\ref{sec:message-status}),
(d) determine the set of users among which each message
has been exchanged (Sec.~\ref{sec:comm-parties}) and, finally, (e)
deal with deleted messages (Sec.~\ref{sec:message-deletion}).
\subsubsection{The structure of the chat database}
\label{sec:chat-db}
The  \textsf{msgstore.db} database contains the following three tables:
\begin{itemize}
\item \textsf{messages}, 
that contains a record for each
message that has been sent or received by the user.
To ease understanding, we classify the fields of these records in two
distinct categories: those storing attributes of the message
(listed in Table~\ref{tab:messages-part1}), and 
those storing the contents of the message and the corresponding metadata
(listed in Table~\ref{tab:messages-part2});
\item \textsf{chat\_list}, that contains a record for
each  conversation held by the user (a conversation consists into the set of messages
exchanged with a particular contact),
whose fields are described in Table~\ref{tab:chat_list};
\item \textsf{sqlite\_sequence}, that stores housekeeping data
used internally by WhatsApp Messenger, whose structure is not reported
here since it does not have any evidentiary value.
\end{itemize}

\begin{table}[hbtp]
\begin{center}
\begin{small}

    \begin{tabularx}{\linewidth}{|l|X|}
        \hline
        \textbf{Field name}  & \textbf{Meaning} \\ \hline\hline
         \_id			 & record sequence number \\ \hline
        key\_remote\_jid & WhatsApp ID of the communication partner  \\ \hline
        key\_id			 & unique message identifier\\ \hline
        key\_from\_me	 & message direction: '0'=incoming, '1'=outgoing\\ \hline
        status			 & message status: '0'=received, '4'=waiting on the central server, '5'=received by the destination, '6'=control message \\ \hline
        timestamp		&  time of send if \textsf{key\_from\_me}='1', record insertion time otherwise (taken from the local device clock, and encoded as a 13-digits millisecond Unix epoch time)\\ \hline
        received\_timestamp	& time of receipt (taken from the local device clock, and encoded as a 13-digits millisecond Unix epoch time) if \textsf{key\_from\_me}='0', '-1' otherwise\\ \hline
        receipt\_server\_timestamp	& time of receipt of the central server ack (taken from the local device clock, and encoded as a 13-digits millisecond Unix epoch time) if \textsf{key\_from\_me}='1', '-1' otherwise \\ \hline 
        receipt\_device\_timestamp		& time of receipt of the recipient ack (taken from the local device clock, and encoded as a 13-digits millisecond Unix epoch time) if \textsf{key\_from\_me}='1', '-1' otherwise \\ \hline   
        
        send\_timestamp	&	unused (always set to '-1')\\ \hline 
        needs\_push	     & '2' if broadcast message, '0' otherwise \\ \hline
        recipient\_count	& number of recipients (broadcast message)\\ \hline 
        remote\_resource	& ID of the sender (only for group chat messages) \\ \hline  	   
         \hline
\end{tabularx} 
\end{small}
\end{center}
\caption{Structure of the \textit{messages} table: fields storing message attributes.}
\label{tab:messages-part1}
\end{table}

\begin{table}[hbtp]
\begin{center}
\begin{small}
    \begin{tabularx}{\linewidth}{|l|X|}
        \hline
        \textbf{Field name}  & \textbf{Meaning} \\ \hline\hline
        media\_wa\_type &  message type: '0'=text, '1'=image, '2'=audio, '3'=video, '4'=contact card, '5'=geo position)\\ \hline
        data			&  message content when \textsf{media\_wa\_type} = '0'\\ \hline
        raw\_data		&  thumbnail of the transmitted file when \textsf{media\_wa\_type}=\{'1','3'\} \\ \hline	
        media\_hash		&  base64-encoded SHA-256 hash of the transmitted file (when \textsf{media\_wa\_type}=\{'1','2','3'\}) \\  \hline 
        media\_url		&  URL of the transmitted file (when \textsf{media\_wa\_type}=\{'1','2','3'\})  \\ \hline 
        media\_mime\_type	&  MIME type of the transmitted file (when \textsf{media\_wa\_type}=\{'1','2','3'\}) \\ \hline 
        media\_size	    &	size of the transmitted file (when \textsf{media\_wa\_type}=\{'1','2','3'\})  \\ \hline 
        media\_name	    & name of transmitted file (when \textsf{media\_wa\_type}=\{'1','2','3'\}) \\ \hline
        media\_duration & duration in sec. of the transmitted file (when \textsf{media\_wa\_type}=\{'1','2','3'\})  \\ \hline
        latitude		&	latitude of the message sender (when \textsf{media\_wa\_type}='5') \\ \hline
        longitude		&   longitude of the message sender (when \textsf{media\_wa\_type}='5') \\\hline
        thumb\_image	&   housekeeping information (no evidentiary value)\\ \hline 
\end{tabularx} 
\end{small}
\end{center}
\caption{Structure of the \textit{messages} table: fields storing information concerning message contents.}
\label{tab:messages-part2}
\end{table}

\begin{table}[hbtp]
\begin{center}
\begin{small}

    \begin{tabularx}{\linewidth}{|l|X|}
        \hline
        \textbf{Field name}  & \textbf{Meaning} \\ \hline\hline
         \_id			 & sequence number of the record \\ \hline
        key\_remote\_jid & WhatsApp ID of the communication partner  \\ \hline
        message\_table\_id & sequence number of record in the \textsf{messages} table that corresponds 
        to the last message of the conversation\\ \hline \hline
\end{tabularx} 
\end{small}
\end{center}
\caption{Structure of the \textit{chat\_list} table.}
\label{tab:chat_list}
\end{table}

As reported in \cite{thakur-master-thesis},
WhatsApp Messenger usually generates various backup copies
of the \textsf{msgstore.db} database, that are stored in the directory
listed in Table~\ref{tab:wa-artifacts-location} row no. 3.
These backups are full copies of the \textsf{msgstore.db}
database, and are not kept synchronized with it.
Therefore, they are particularly important from an investigative standpoint, 
since they 
may store messages that have been deleted from the main chat database. 
Backups are encrypted with the AES 192 algorithm, but they
can be easily decrypted since, as discussed in~\cite{wa:encryption},
the same encryption key 
(namely, \texttt{346a23652a46392b4d73257c67317e352e3372482177652c})
is used on all devices.

\subsubsection{Reconstruction of the chat history}
\label{sec:chat-history}
To reconstruct the chronology of the messages exchanged by the user, the
records stored in the \textsf{messages} table must be extracted and decoded
as discussed below.

\begin{figure}[htbp]
\centering
\includegraphics[scale=0.6]{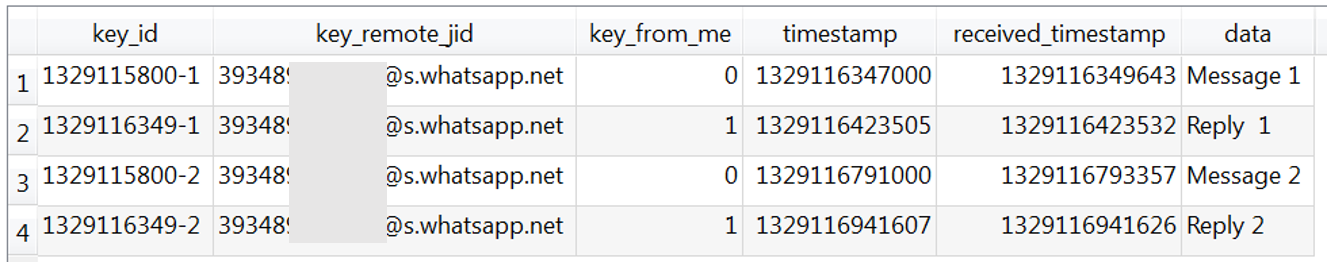}
\caption{Reconstruction of the chat history. Phone numbers have been grayed out to ensure the privacy of the owner.}
\label{fig:chat_history_example}
\end{figure}

To elucidate, let us consider Fig.~\ref{fig:chat_history_example}, 
that shows four records corresponding to
a conversation between the device owner and the user
\textsf{39348xxxxxxx} (actually, only the fields listed in Table~\ref{tab:messages-part1} 
are displayed).

By examining these records, we note that (a) all the messages have been exchanged with the same 
contact \textsf{39348xxxxxxx} (they all store the same WhatsApp ID in the \textsf{key\_remote\_id} field), 
(b) the conversation has been started by that contact (\textsf{key\_from\_me} = '\textsl{0}' in record no.1) 
with a textual message whose content was ``\textsl{Message 1}'' (field \textsf{data}) on 
Feb. 13th, 2012 06:59:09 (field \textsf{received\_timestamp}),
and (c) the device owner replied at 07:00:23 of the same day (field \textsf{timestamp})
with the message corresponding to record no. 2
(\textsf{key\_from\_me}='\textsl{1}') with content ``\textsl{Reply 1}'' (field \textsf{data}).
The conversation then continued with another message-reply exchange.

From these records, we also note that each message carries its own unique identifier
in the \textsf{key\_id} field: this value, set by the sender, is obtained by
concatenating the timestamp corresponding to the last start time of WhatsApp
Messenger (on the sender's device) with a progressive number (indicating the number of
messages sent from that moment), and is used also by the recipient to
denote that message.
Therefore, by using this value, it is possible to correlate the records of
the sender's and recipient's databases corresponding to the same message.

\subsubsection{Extracting the contents of a message}
\label{sec:conversation-content}
In addition to plain text messages, WhatsApp allows its users to exchange messages containing data of various
types, namely multimedia files (storing images, audio, and video), contact cards, 
and geo-location information.
The type of data transmitted with a message is indicated (as reported in
Table~\ref{tab:messages-part2}) by the \textsf{media\_wa\_type} field, while the information
concerning message content is spread, for non-textual messages, over
several fields (depending on the specific data type).
As a matter of fact, while the content of textual messages (\textsf{media\_wa\_type='0'})
is stored in the \textsf{data} field, for the other types of contents the situation is more 
involved, as discussed below.

\paragraph{Multimedia files} 
When the user sends a multimedia file, 
several activities take place automatically (i.e., without informing the involved
users).
First, WhatsApp Messenger copies the file into the 
folder listed in  Table~\ref{tab:wa-artifacts-location}, row 8.
Then, it uploads the file to 
the WhatsApp server, that sends back the URL of the corresponding location.
Finally, the sender sends to the recipient a message
containing this URL and,
upon receiving this message, the recipient sends an acknowledgment back to
the sender.

\begin{figure}[hbtp]
\centering
\includegraphics[scale=0.6]{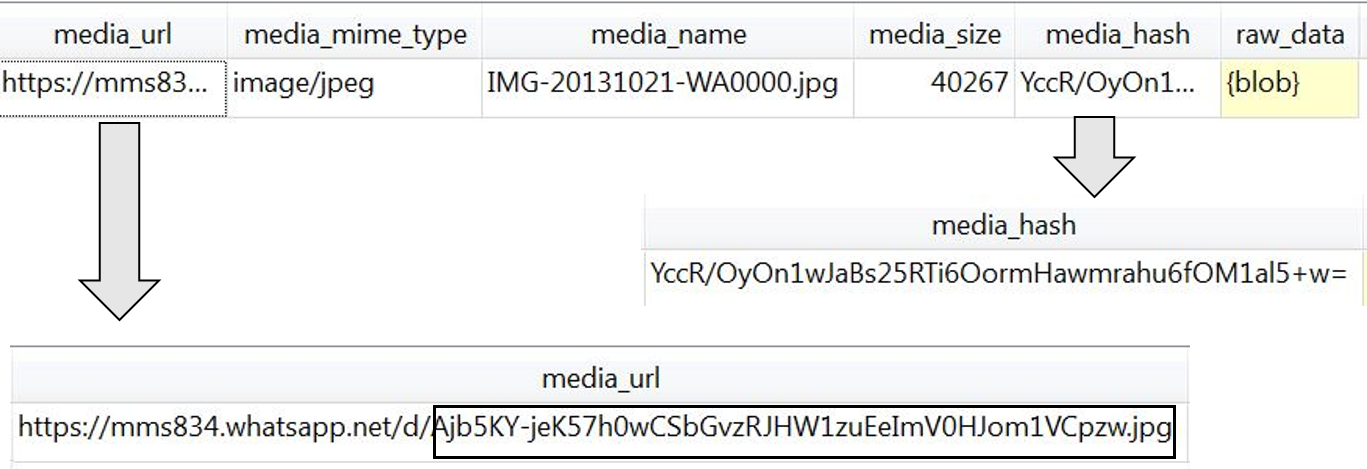}
\caption{Multimedia file exchange: sender side}
\label{fig:multimedia-sender}
\end{figure}

When these steps have been completed, the sender stores into
his/her \textsf{messages} table a record like the
one shown in Fig.~\ref{fig:multimedia-sender} (where we show only
the fields related to message contents that are listed in
Table~\ref{tab:messages-part2}).
As can be seen from the above figure,
the type of the file is indicated 
(besides the \textsf{wa\_media\_type} field, not shown in the figure)
by the \textsf{media\_mime\_type} field ('\textsl{image/jpeg}' in the example).
Its name is instead stored in the
\textsf{media\_name} field (\textsl{IMG-20131021-WA0000.jpg} in the example),
its size in bytes by \textsf{media\_size} (\textsl{40267} in the example), and
its thumbnail in the \textsf{raw\_data} field (as a \textsl{blob}, i.e. a binary large object).
Furthermore, 
the \textsf{media\_url} field stores the URL of the 
location on the central server where the file has been temporarily stored,
whose last part (highlighted in Fig.~\ref{fig:multimedia-sender} by framing it)
corresponds to the name given by the server to that file.
Finally, the base64-encoded SHA-256 hash of the transmitted file is stored
in the \textsf{media\_hash} field.

\begin{figure}[hbtp]
\centering
\includegraphics[scale=0.6]{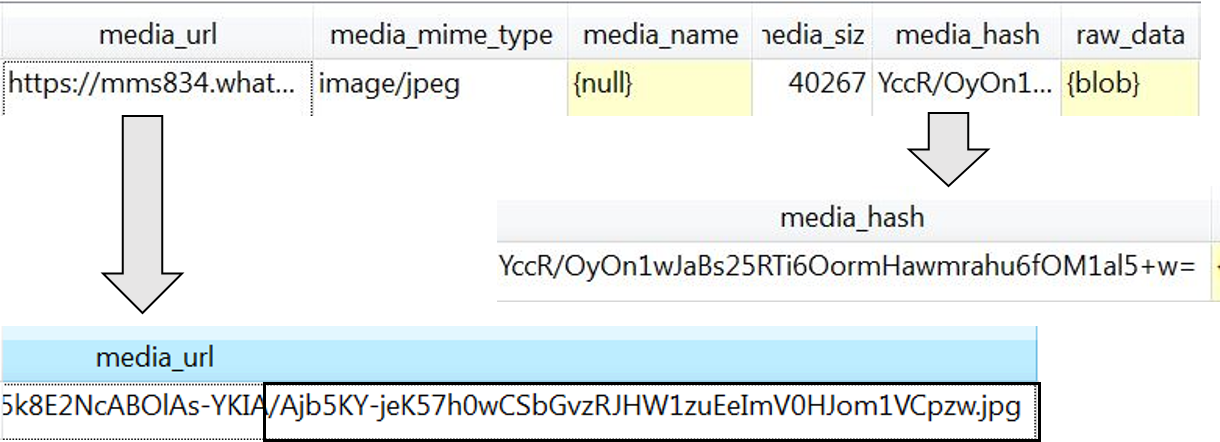}
\caption{Multimedia file exchange: recipient side}
\label{fig:multimedia-receiver}
\end{figure}

On the recipient side, after message reception, the transmitted thumbnail of the
file is displayed by
WhatsApp Messenger; the actual file is instead downloaded 
at a later time only if the recipient explicitly requests it.
Upon message reception, the recipient stores in his/her
\textsf{messages} table a record like the one shown in
Fig.~\ref{fig:multimedia-receiver}.
From this figure, we see that most fields are identical to those
stored by the sender (in particular, \textsf{wa\_media\_type}, \textsf{media\_mime\_type},
\textsf{media\_size}, \textsf{raw\_data}, 
and \textsf{media\_hash}).
Conversely, the contents of \textsf{media\_url} is different,
except for the name given to the file by the server (highlighted in Fig.~\ref{fig:multimedia-receiver} by framing it).

Unlike the sender, however, the \textsf{media\_name} field is empty, so the local
name given by WhatsApp Messenger to that file is unknown. The file can be however identified
by comparing the SHA-256 hash stored into the corresponding record 
with that of all the files that have been received (that are stored
in the folder reported in Table~\ref{tab:wa-artifacts-location}, row 7.

Finally, we note that the file sent by the sender and the one received
by the recipient can be correlated by comparing both the file name
given by the WhatsApp server and the SHA-256 hash to these files
(that are stored, as discussed above, in the \textsf{media\_url} and
\textsf{media\_hash} fields of the corresponding records).

\begin{figure}[hbtp]
\centering
\includegraphics[scale=0.7]{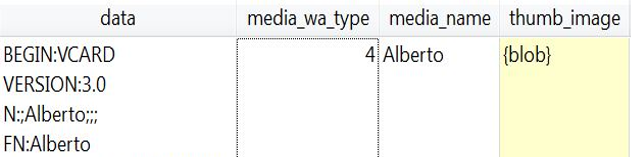}
\caption{Fields containing contact card information.}
\label{fig:contact-card}
\end{figure}

\paragraph{Contact cards} 
Messages carrying contacts cards (extracted from the phonebook of the
sender) correspond -- both on the sender and on the recipient side --
to \textsf{messages} record that store the 
transmitted information (in \textsf{VCARD} format)
into the \textsf{data} field, and 
the name given by the sender to that contact in the \textsf{media\_name} field.
An example of such a record is shown in Fig.~\ref{fig:contact-card}.

\begin{figure}[hbtp]
\centering
\includegraphics[scale=0.6]{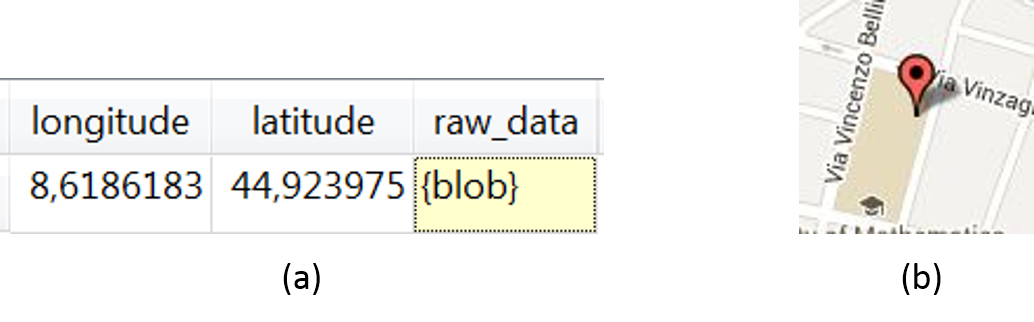}
\caption{Geo-location message:(a) data stored in the database, (b) Google map extracted from the \textsf{raw\_data} field.}
\label{fig:sender-geo}
\end{figure}

\paragraph{Geolocation coordinates} 
WhatsApp Messenger enables users to send the geographic coordinates
of their current location, that are obtained from the Android Location
Services running on the device.
Messages carrying geographic coordinates correspond --both on the sender
and on the recipient side -- to \textsf{messages}
records that store the latitude and the longitude values into the \textsf{latitude}, \textsf{longitude} fields, and a JPEG thumbnail of the
Google Map displaying the above coordinates in the \textsf{raw\_data}
field.
An example of such a record is shown in Fig.~\ref{fig:sender-geo}.

\subsubsection{Determining the state of the message}
\label{sec:message-status}
In WhatsApp, messages are not exchanged directly among
communicating users, but they are first sent to the central server,
that forwards them to the respective recipients if they are on-line,
and stores them locally until they can be delivered otherwise.
This implies that the presence of a record in the
\textsf{messages} table does not necessarily mean that an outgoing message 
has been actually delivered to its recipients.
As a matter of fact, after the user has pushed the ``send'' button of
WhatsApp Messenger, the message can be in one of the following three states: (a)
waiting on the local device to be transmitted to the central server, or
(b) stored on the central server but  waiting to be transmitted to its recipient(s), 
or (c) delivered to its recipient(s).

The ability to distinguish the various states of a message may be
crucial in an investigation where it must be ascertained whether 
a message has been actually delivered or not to its destinations, and when
such a delivery has taken place.

The current state of a message, as well as the times of its state
changes, can be understood by correlating the values contained in
several fields of the corresponding record of the
sender database~\footnote{For a recipient, a message can be only
in the received state, corresponding to \textsf{status}='0'},
namely \textsf{status}, \textsf{timestamp}, \textsf{received\_timestamp},
\textsf{receipt\_server\_timestamp}, and \textsf{receipt\_device\_timestamp}.

\begin{figure}[hbtp]
\centering
\includegraphics[scale=0.7]{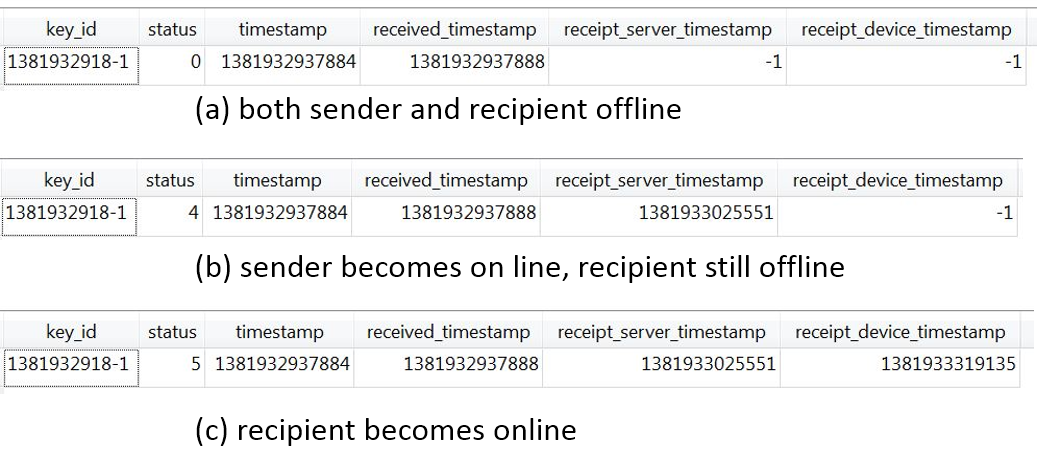}
\caption{Sender side: record updates for a message while in transit.}
\label{fig:message-status}
\end{figure}

To explain, let us consider a scenario in which a user
sends a message when both him/her and the recipient are off-line
(Fig.~\ref{fig:message-status}(a)),
then the sender gets reconnected to the network while the recipient is 
still offline (Fig.~\ref{fig:message-status}(b)), and then, finally, also
the recipient gets connected (Fig.~\ref{fig:message-status}(c)).

When the message is sent, a record is stored in the \textsf{messages}
table of the sender, even if the central server is unreachable.
In this case, as shown in Fig.~\ref{fig:message-status}(a), in this
record we have that \textsf{status}='0', \textsf{timestamp}='\textsl{x}', and 
\textsf{received\_timestamp}='\textsl{y}',
where '\textsl{x}' and '\textsl{y}' correspond to when the user has sent the message
and when the record has been added to the chat database, respectively.

\emph{Thus, a record such that \textsf{key\_from\_me}='1' and \textsf{status}='0'
corresponds to a message that has not been delivered to the central server yet}.

Later, when the sender returns on-line, 
the message is forwarded to the central server that replies with an ack.
When this ack is received, the sender updates the corresponding record
as shown in Fig.~\ref{fig:message-status}(b)
by setting \textsf{status}='4', and the value of
\textsf{receipt\_server\_timestamp} to the reception time of the ack.

\emph{Thus, a record such that \textsf{key\_from\_me}='1' and \textsf{status}='4'
corresponds to a message that has been delivered the central server,
but not yet to its destination(s)}.

Finally, when the recipient returns on line, it receives the message from
the central server, and sends an ack to the sender.
Upon receiving this ack, the sender
updates again the record corresponding to that message (as shown in Fig.~\ref{fig:message-status}(c))
by setting \textsf{status}='5', and the value of \textsf{receipt\_device\_timestamp} to the
reception time of the ack.

\emph{Thus, a record such that \textsf{key\_from\_me}='1' and \textsf{status}='5'
corresponds to a message that has been delivered to its destination}.

From the above discussion, it follows that the times of the state changes of 
a message can be tracked by means of
the values stored in the various timestamp fields of the corresponding record.
For instance, in the case in Fig.~\ref{fig:message-status}(c), we can deduce that
the message has been generated on Oct. 16th, 2013 14:15:37.884 (\textsf{timestamp} field), 
has been waited
to be transmitted to the central server until 14:17:05.551 of the same day (\textsf{receipt\_server\_timestamp} field),
and has been finally delivered to its recipient at 14:21:59.135 (\textsf{receipt\_device\_timestamp} field).

\subsubsection{Determining the partners of a message}
\label{sec:comm-parties}
In addition to user-to-user communication, 
WhatsApp provides its users with two forms of collective
communications, namely:
\begin{itemize}
\item
\textsl{broadcast} (i.e, one-to-many) communication, whereby a user 
(the \textsl{source user}) sends the
same message to a group of other users (the \textsl{destination users}) that are not aware of each
other and whose possible replies are sent to the source user only;
\item \textsl{group chats}, providing a many-to-many communication service, 
whereby each message sent by any user belonging to a chat is received by all the users 
belonging to that chat.
\end{itemize}

While the WhatsApp ID of the communication partner in a 
user-to-user communication is easily retrieved from the
\textsf{key\_remote\_jid} field, to determine the set of users
involved into a broadcast or a group chat message
various fields have to be correlated, as discussed below.

\begin{figure}[hbtp]
\centering
\includegraphics[scale=0.6]{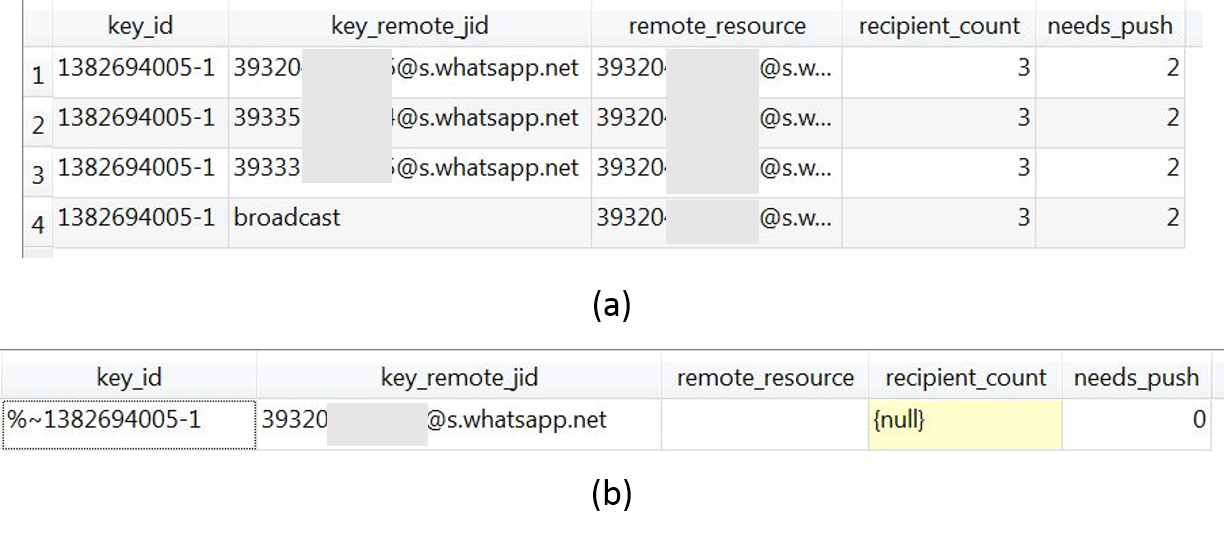}
\caption{Records generated for a broadcast message sent to three recipients on: (a) the sender, (b) one of the recipients. Only the fields that contribute to
the identification of the partners are displayed.}
\label{fig:sender-bcast}
\end{figure}

\paragraph{Broadcast messages}
When a user sends a broadcast message, a distinct record is
created in his/her \textsf{messages} table for each one of the
recipients, plus one for itself, as reported in Fig.~\ref{fig:sender-bcast}(a),
that shows the records generated by a broadcast message sent to
users \textsl{39320xxxxxxx}, \textsl{39335xxxxxxx}, and 
\textsl{39333xxxxxxx}. 

As shown in this figure, all the records corresponding to the same broadcast message
have the same message identifier (stored in the \textsf{key\_id} field),
so they can be easily identified.
Each one of these records stores in the \textsf{key\_remote\_jid}
field the WhatsApp ID of the recipient
(the sender uses the keyword \textsl{broadcast} to denote
itself as a recipient), while
the \textsf{remote\_resource} and the \textsf{recipient\_count} fields
store the WhatsApp IDs of the set of destinations and how many they are,
respectively (field \textsf{needs\_push}
instead always stores the value '\textsl{2}').

The situation on each one of the destinations is instead 
different (Fig.~\ref{fig:sender-bcast}(b)), since each one of
them stores, in his/her \textsf{messages} table, only a single record
that is generated when it receives the broadcast message.
This record can be distinguished from those corresponding to
non-broadcast messages by looking at the value stored in its
\textsf{key\_id} field, that consists in the
concatenation of the \textsl{\%$\sim$} characters  with the
message identifier set by the sender.

\paragraph{Group chat communication}
When a message is sent within a group chat,
a  record is generated in the \textsf{messages} table of
all the members of that group (including the sender).
Each one of these records stores, in the \textsf{key\_remote\_jid} field,
the identifier of the group (the \textsl{group\_id}), a string formatted as \textsl{\{creator's phone number\}-\{creation time\}@g.us} (where \textsl{creation time} is encoded as a Unix epoch time). 

To illustrate, consider a group chat consisting of three members,
namely  \textsl{3933xxxxxxx},
\textsl{3936xxxxxxx}, and \textsl{3932xxxxxx}
(in the following denoted as \textsl{A}, \textsl{B}, and \textsl{C}, respectively, for brevity),
where each user sends in turn  to the group a message
with textual content '\textsl{Message from X}' (where 'X' is
the name of the user).

\begin{figure}[hbtp]
\centering
\includegraphics[scale=0.55]{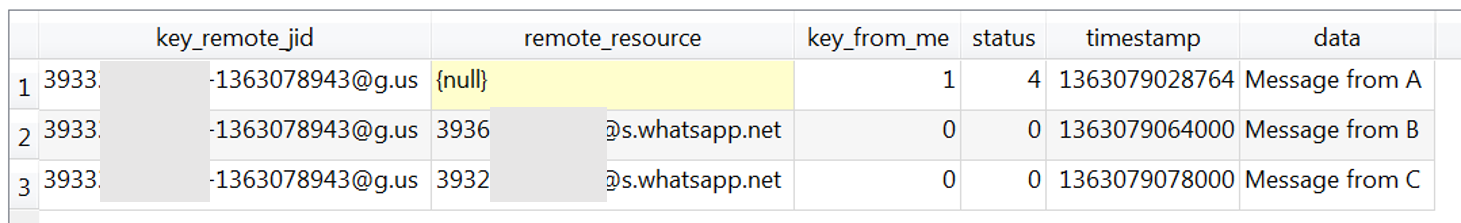}
\caption{Records correspoding to three messages exchanged within a group chat.}
\label{fig:group_chat}
\end{figure}

Let us focus on the
records stored in the \textsf{messages} table of user
\textsl{A} at the end of this exchange, that are shown
in Fig.~\ref{fig:group_chat} (the situation for the
other users is identical).

As can be seen from this figure, all these records store the same group\_id
\textsl{3933xxxxxxx-1363078943@g.us}
in the \textsf{key\_remote\_id} field.
From this value, we can determine the
creator of the group (user \textsl{A}) and the date and hour of group creation
(March 12, 2013 at 09:02:23).
Furthermore, the WhatsApp ID of the message originator is stored into 
the \textsf{remote\_resource} field, while the time of message receipt is stored into the
\textsf{timestamp} field.
Note that \textsl{A} stores also the record corresponding to the message
that (s)he has sent to the group (record no. 1 in the figure).
Records like this one can be easily identified by looking at
the contents of their  \textsf{status} and \textsf{remote\_resource}
fields, that store the values '\textsl{4}' and '\textsl{null}', respectively.

Note also that the set of recipients, i.e. of
the set of group members at the time of the sending, is not stored
anywhere on the record.
However, it can be indirectly determined by examining the
records corresponding to the \textsl{control messages}
that are automatically exchanged by the various group members
each time a user joins or leaves the group.
These messages, also stored in the \textsf{messages} table, 
always contain  the  value '\textsl{6}' 
in the \textsf{status} field, and encode in the \textsf{media\_size}
field the specific operation corresponding to the message
(in particular, the values '\textsl{1}', '\textsl{4}', and
'\textsl{5}' indicate group creation, join, and leave,
respectively).

To illustrate, let us consider a scenario in which 
user \textsl{39320xxxxxxx} (\textsl{D}, for brevity)
creates a group on Nov. 11, 2013 at 16:24:05, and
immediately adds user  
\textsl{39335xxxxxxx} (\textsl{E}, for brevity)
to the group.
Then, \textsl{D} adds user  \textsl{39333xxxxxxx}
(\textsl{F}, for brevity)
on Nov. 12, 2013 at 10:40:48.

\begin{figure}[hbtp]
\centering
\includegraphics[scale=0.6]{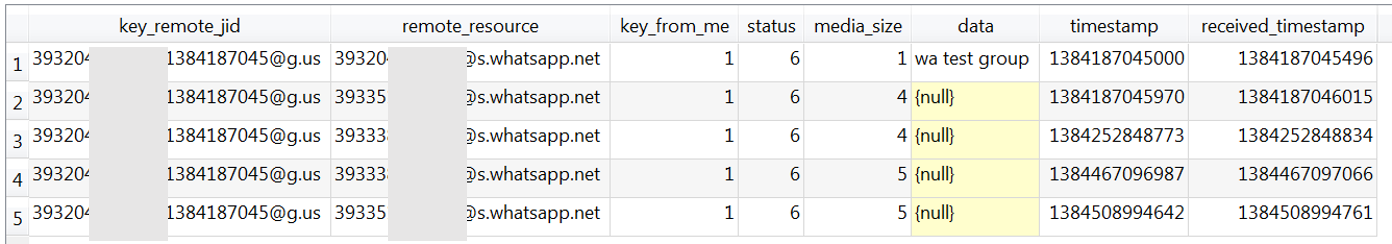}
\caption{Group management records stored in the msgstore database of user \textsl{D}.
For other users we have the same situation, with the exception of record no. 1.
}
\label{fig:group_creation}
\end{figure}

The records generated by these operations in the chat database
of user \textsl{D} are shown in Fig.~\ref{fig:group_creation} 
(the same situation occurs on all the other group members).

Group creation corresponds to record no. 1,
as can be seen from \textsf{status}='\textsl{6}' and
\textsf{media\_size}='\textsl{1}'.
The time of group creation can be ascertained
(besides from the group\_id) from the value stored in the
\textsf{timestamp} field, while the field \textsf{data} stores the name given to the
group (\textsl{wa test group}).

The addition of user \textsl{E} corresponds instead to record
no. 2: the specific operation (join) and the identity of the user joining
the group (\textsl{E}) can be deduced from fields \textsf{media\_size}
and \textsf{remote\_resource} field, while the time of occurrence
is stored in the \textsf{timestamp} field.
A similar situation occurs with the addition of user \textsl{F},
whose control message corresponds  to record no. 3.

Now, suppose that at a later time, namely on Nov. 14, 2013 at 22:11:36,
user \textsl{F} leaves the group. 
This operation corresponds to record no. 4 in Fig.~\ref{fig:group_creation}
(\textsf{media\_size}='\textsl{5}' indicates a group leave),
that reports the identity of the user leaving the group and the time
of leave in the \textsf{remote\_resource} and the \textsf{timestamp} field,
respectively.
Finally, when user \textsl{E} leaves the group on Nov. 15, 2013 at
09:49:54, record no. 5 is added to the \textsf{messages} table.

\begin{figure}[hbtp]
\centering
\includegraphics[scale=0.5]{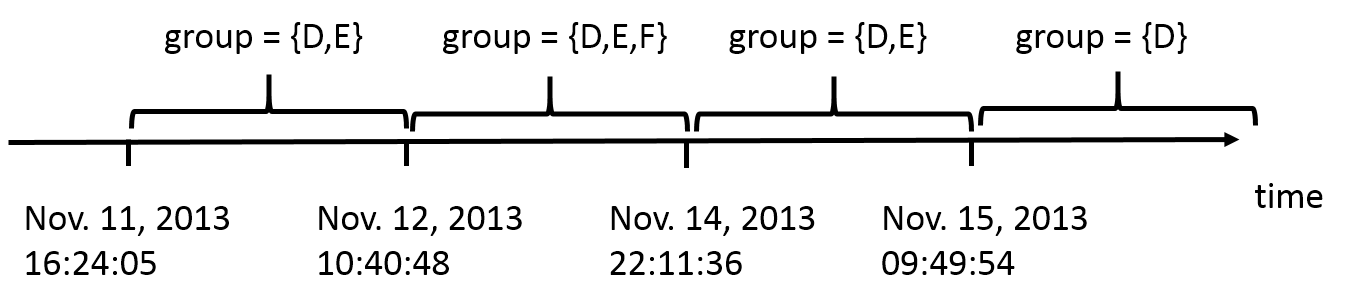}
\caption{Timeline of group composition variations.}
\label{fig:group_timeline}
\end{figure}

By using the information discussed above, the composition of the
group over time can be reconstructed by chronologically sorting
the various control messages corresponding to join (\textsf{status}='6'
and \textsf{media\_size}='4') and leave (\textsf{status}='6'
and \textsf{media\_size}='5') of a given group 
(identified by the contents of the \textsf{key\_remote\_jid} field),
as shown in Fig.~\ref{fig:group_timeline}.

From this information, it can be inferred whether a user belonged or
not to a group when a specific message was sent to that group.

\subsubsection{Dealing with deleted messages}
\label{sec:message-deletion}
In WhatsApp Messenger, the user may delete the records
stored in the \textsf{msgstore.db} database in two different ways,
namely:
\begin{itemize}
\item deletion of an individual message: in this case,
the corresponding record is deleted from the \textsf{messages} table;
\item deletion of all the records belonging to a one-to-one,
broadcast, or group chat conversation:
in this case, all the records corresponding
to the messages exchanged in that conversation are deleted from the
\textsf{messages} table, as well as the
record of the \textsf{chat\_list} table corresponding to that conversation.
\end{itemize}

As discussed before, it is sometimes possible to recover deleted
SQLite records, and in these cases the analysis techniques discussed
in the previous sections can be applied.

However, when such a recovery is not feasible, it may be still possible
to determine many of the information regarding a deleted message
by analyzing the log files generated by WhatsApp Messenger.
In particular, as discussed below, it is possible to determine
which messages have been deleted and when, when a deleted message has been sent 
or received and its state, as well as the users involved in the
conversation.
The same holds true for group control messages, so the
analysis of log files also makes it possible to track the evolution of each group
over time.
In other words, only the contents of a deleted message cannot
be recovered anymore.

\begin{figure}[hbtp]
\centering
\includegraphics[scale=0.8]{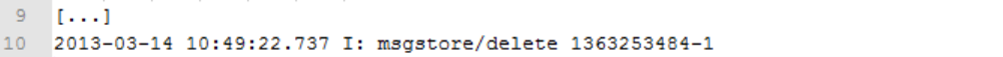}
\caption{Events logged when a message is deleted.}
\label{fig:message_deleted}
\end{figure}

\paragraph{Finding deleted messages and their deletion times}
When a message is deleted, WhatsApp Messenger records into the
log file an event like the one shown in
Fig.~\ref{fig:message_deleted}, that indicates both the
type of operation (\textsf{msgstore/delete}) and the identifier
of the deleted message (\textsl{1363253484-1}), as well as
the time of deletion (March 14, 2013 at 10:49:22).

\paragraph{Determining when a deleted message has been exchanged, and its state}
Each time a user-to-user, broadcast, or group chat
message is sent/received, WhatsApp Messenger logs
the time of the send/receive operation, the identifiers of the
involved users, and the identifier of the message.
Therefore, by searching into the log file the events
corresponding to exchanges of deleted messages, it is possible to ascertain 
when those messages have been sent or received.

\begin{figure}[hbtp]
\centering
\includegraphics[scale=0.8]{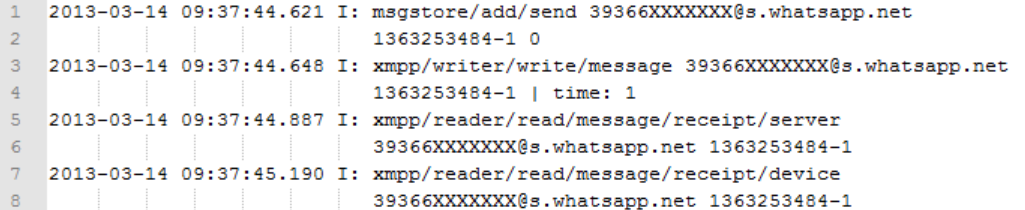}
\caption{Events logged when a message is sent.}
\label{fig:message_exchange}
\end{figure}

For instance, 
the event reported in Fig.~\ref{fig:message_exchange} line 1 corresponds to the sending of 
the deleted message identified
by \textsl{1363253484-1} to user \textsl{39366xxxxxxx} on March 14, 2013 at 09:37:44.

Finally, WhatsApp Messenger
logs also the events corresponding to reception of the acknowledgment messages
sent back by the central server (line no. 5) and by the recipient (line no. 7),
from which it is possible to determine the state of a message, as well
as the times of its state changes.

\paragraph{Temporal evolution of group chat composition}
The evolution of the composition of a group chat can be
tracked over time by examining the events, logged by WhatsApp 
Messenger, corresponding to the exchange of the various
control messages discussed in Sec.~\ref{sec:comm-parties}.

\begin{figure}[hbtp]
\centering
\includegraphics[scale=0.7]{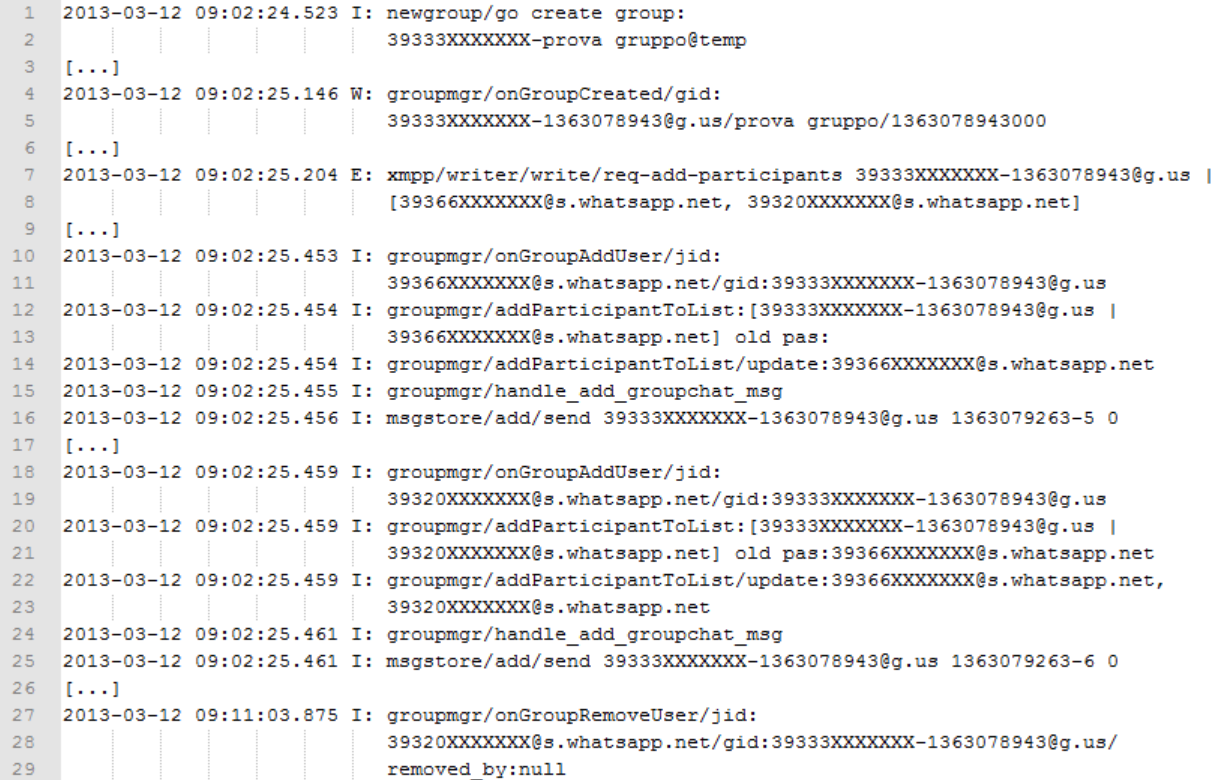}
\caption{Events corresponding to group operations.}
\label{fig:group_creation_message}
\end{figure}

The events corresponding to the creation of a group and to
the addition of two users are shown in  Fig.~\ref{fig:group_creation_message}
(that reports an excerpt of the log of the group creator).

The creation of the group gives rise to the events reported in lines
no. 1 and 4, from which we can obtain the group name and creation time.
The request to add to the group users
\textsl{39366xxxxxxx} and \textsl{39320xxxxxxx}
corresponds to line 7, while the addition of the former and the latter
user corresponds to lines 10--14 and 18--22. 
Finally, the leave of a user corresponds to the event logged on line no. 27.

\subsection{Analysis of settings and preferences}
\label{sec:settings-and-preferences}
WhatsApp Messenger stores various information of potential
evidentiary value in several files, located in 
the directories listed in Table~\ref{tab:wa-artifacts-location} row no. 9.

In particular, the file \textsf{me} stores (as ASCII text) the phone number 
registered with WhatsApp (i.e., the number used to create the corresponding
WhatsApp ID).
The relevance of this information derives from the fact that
the SIM card currently used with the smartphone may not be the
one used to register the user with WhatsApp: it is indeed possible
to replace the latter SIM card with a new one, and to use 
the WhatsApp ID corresponding to the phone number of the
old SIM card.
Thus, a user \textsl{A} may impersonate a different user \textsl{B}, 
as long as \textsl{A} has used \textsl{B}'s SIM card during registration,
or (s)he is using \textsl{B}'s smartphone with a different SIM card.
By comparing the phone number of the SIM inserted into a smartphone
with the phone number stored in the \textsf{me} file, it is possible
to determine whether this is the case or not.

Furthermore, the file \textsf{me.jpg} stores the currently-used avatar picture 
of the user. Given that the avatar pictures of all contacts are downloaded
locally by WhatsApp Messenger (as discussed in Sec.~\ref{sec:getting-contacts}),
the \textsf{me.jpg} file can be used to understand that the 
user of the device under examination has been in contact with another user 
even if the latter one has deleted from its contacts database the
record corresponding to the former one.
As a matter of fact, the deletion of a record from the contacts database
does not cause the deletion of his/her downloaded avatar picture.

\section{Conclusions}
\label{sec:conclusions}
In this paper we have discussed the forensic analysis of the artifacts
left by WhatsApp Messenger on Android smartphones, and we have shown
how these artifacts can provide many information of evidentiary value.
In particular, we have shown how to interpret the data stored into the contacts
and chat databases in order to reconstruct the list of
contacts and the chronology of the messages that have been exchanged
by users.

More importantly, we have also shown the importance of correlating among them the
artifacts generated by WhatsApp Messenger in order to gather information that cannot be
inferred by examining them in isolation.
As a matter of fact, while the analysis of the contacts database 
makes it possible to reconstruct the list of contacts, the correlation
with the events stored into the log files maintained by WhatsApp Messenger
allows the investigator to infer also when a specific contact has been
added, or to recover deleted contacts and their time of deletion.
Similarly, the correlation of the contents of the chat database with
the information stored into the log files allows the investigator to
determine which messages have been deleted, when these messages have been
exchanged, and the users that exchanged them.

The results reported in this paper have a two-fold value.
On the one hand, they provide analysts with the full 
picture concerning the decoding, interpretation,
and correlation of WhatsApp Messenger artifacts on
Android devices.
On the other hand, they represent a benchmark against which
the ability of extraction tools for smartphone to retrieve
all the WhatsApp Messenger artifacts can be assessed.

It is however worth to point out that the results discussed in this paper
apply to Android only: as a matter of fact, there is evidence~\cite{tso-iphone} 
showing that on different platforms (e.g., iOS) WhatsApp Messenger  
produces artifacts that differ either in the information they store, or
in their format, or in both.
We leave the analysis of WhatsApp Messenger for other smartphone
platforms as future work.

\bibliographystyle{plain}
\bibliography{biblio}

\begin{thebibliography}{10}

\bibitem{ftk-imager}
{AccessData Corporation}.
\newblock {FTK Imager}, 2013.
\newblock Available at http://www.accessdata.com/support/product-downloads.

\bibitem{social-im-forensics}
N.B.~Al Barghuthi and H.~Said.
\newblock {Social Networks IM Forensics: Encryption Analysis}.
\newblock {\em Journal of Communications}, 8(11), Nov. 2013.

\bibitem{bellovin-wiretapping}
Steven~M. Bellovin, Matt Blaze, Sandy Clark, and Susan Landau.
\newblock {Lawful Hacking: Using Existing Vulnerabilities for Wiretapping on
  the Internet}.
\newblock In {\em Proc. of Privacy Legal Scholars Conference}, June 2013.
\newblock Available at http://dx.doi.org/10.2139/ssrn.2312107.

\bibitem{ufed}
{Cellebrite LTD.}
\newblock {Cellebrite Android Forensics}, 2013.
\newblock Available at
  http://www.cellebrite.com/mobile-forensics/capabilities/android-forensics.

\bibitem{wa:encryption}
D.~Cortjens, A.~Spruyt, and W.F.C. Wieringa.
\newblock {WhatsApp Database Encryption Project Report}.
\newblock Technical report, 2011.
\newblock Available at
  https://www.os3.nl/\_media/2011-2012/students/ssn\_project\_report.pdf.

\bibitem{notepad++}
Don Ho.
\newblock {Notepad++ Home}, 2013.
\newblock Available at http://notepad-plus-plus.org.

\bibitem{hoog-book}
Andrew Hoog.
\newblock {\em {Android Forensics: Investigation, Analysis and Mobile Security
  for Google Android}}.
\newblock Elsevier - Syngress, 2011.

\bibitem{iforensics}
MohammadIftekhar Husain and Ramalingam Sridhar.
\newblock {iForensics: Forensic Analysis of Instant Messaging on Smart Phones}.
\newblock In Sanjay Goel, editor, {\em Digital Forensics and Cyber Crime},
  volume~31 of {\em Lecture Notes of the Institute for Computer Sciences,
  Social Informatics and Telecommunications Engineering}. Springer Berlin
  Heidelberg, 2010.

\bibitem{sqlite-deletion}
S.~Jeon, J.~Bang, K.~Byun, and S.~Lee.
\newblock {A recovery method of deleted records for SQLite database}.
\newblock {\em {Personal and Ubiquotous Computing}}, 16, 2012.

\bibitem{mahajan-ijca-2013}
A.~Mahajan, M.S. Dahiya, and H.P. Sanghvi.
\newblock {Forensic Analysis of Instant Messenger Applications on Android
  Devices}.
\newblock {\em {International Journal of Computer Applications}}, 68(8), April
  2013.

\bibitem{xry}
{Micro Systemation}.
\newblock {XRY}, 2013.
\newblock Available at http://www.msab.com/xry/xry-current-version.

\bibitem{wa:300m}
Nathan Olivarez-Giles.
\newblock Whatsapp adds voice messaging as it hits 300 million monthly active
  users.
\newblock {The Verge}, Aug 2013.
\newblock Available at
  http://www.theverge.com/2013/8/6/4595496/whatsapp-300-million-active-users-voice-messaging-update.

\bibitem{virtualbox}
{Oracle Corp.}
\newblock {Oracle VM Virtual Box}.
\newblock http://www.virtualbox.org, 2013.

\bibitem{oxygen}
{Oxygen Forensics, Inc.}
\newblock {Oxygen Forensics}, 2013.
\newblock Available at http://www.oxygen-forensic.com/en/features/analyst.

\bibitem{oxygen-sqlite}
{Oxygen Forensics, Inc.}
\newblock {SQLite Viewer}, 2013.
\newblock Available at
  http://www.oxygen-forensic.com/en/features/analyst/data-viewers/sqlite-viewer.

\bibitem{foxit-dfrws2011}
Ivo Pooters, Pascal Arends, and Steffen Moorrees.
\newblock {Extracting SQLite records: Carving, parsing and matching}.
\newblock Technical report, {Digital Forensics Research Workshop Challenge},
  2011.
\newblock Available at
  http://sandbox.dfrws.org/2011/fox-it/DFRWS2011\_results/Report/Sqlite\_carving\_extractAndroidData.pdf.

\bibitem{sangiacomo-wa-xtract}
Fabio Sangiacomo and Martina Weidner.
\newblock {WhatsApp Xtract (v. 2.1)}, May 2012.
\newblock Available at https://code.google.com/p/hotoloti/downloads/list.

\bibitem{sqlite3}
{SQLite Consortium}.
\newblock {SQLite Home Page}, 2013.
\newblock Available at http://www.sqlite.org.

\bibitem{thakur-master-thesis}
N.S. Thakur.
\newblock {Forensic Analysis of WhatsApp on Android Smartphones}.
\newblock Master's thesis, University of New Orleans, 2013.
\newblock Paper 1706.

\bibitem{unodc-2013}
{The United Nations Office on Drugs and Crime}.
\newblock {Comprehensive Study on Cybercrime}.
\newblock Technical report, United Nations, Feb. 2013.
\newblock Available at
  http://www.unodc.org/documents/organized-crime/UNODC\_CCPCJ\_EG.4\_2013/CYBERCRIME\_STUDY/
  \_210213.pdf.

\bibitem{tso-iphone}
Yu-Cheng Tso, Shiuh-Jeng Wang, Cheng-Ta Huang, and Wei-Jen Wang.
\newblock {iPhone Social Networking for Evidence Investigations Using iTunes
  Forensics}.
\newblock In {\em Proceedings of the 6th International Conference on Ubiquitous
  Information Management and Communication}, ICUIMC '12, New York, NY, USA,
  2012. ACM.

\bibitem{sqliteman}
Petr Vanek and Kamil Les.
\newblock {Sqlite Databases Made Easy}, 2013.
\newblock Available at http://sqliteman.com.

\bibitem{android-acquisition}
Timothy Vidas, Chengye Zhang, and Nicolas Christin.
\newblock {Towards a General Collection Methodology for Android Devices}.
\newblock {\em Digital Investigation}, 8, Aug 2011.

\bibitem{whatsapp}
{WhatsApp Inc.}
\newblock {WhatsApp}, 2013.
\newblock Available at http://www.whatsapp.com.

\bibitem{wa:400m}
Rolfe Winkler.
\newblock {WhatsApp Hits 400 Million Users, Wants to ‘Stay Independent’}.
\newblock {The Wall Street Journal - Digits}, Oct. 2013.
\newblock Available at
  http://blogs.wsj.com/digits/2013/12/19/whatsapp-hits-400-million-users-wants-to-stay-independent.

\bibitem{youwave}
{YouWave Corp.}
\newblock Youwave home page, 2013.
\newblock Available at http://youwave.com.

\end{thebibliography}

\end{document}